\documentclass[12pt]{article}
\usepackage{amsmath}
\usepackage{cite}
\usepackage{graphicx}

\csname @addtoreset\endcsname{equation}{section}
\newcommand{\eq}{\begin{equation}}
\newcommand{\feq}{\end{equation}}
\newcommand{\eqn}{\begin{eqnarray}}
\newcommand{\feqn}{\end{eqnarray}}
\newcommand{\arr}{\begin{eqnarray*}}
\newcommand{\farr}{\end{eqnarray*}}

\input{tcilatex}

\begin{document}

\begin{titlepage}
\begin{flushright}
HUTP-02/A025\\
CAMS/02-03\\
hep-th/0206153\\
\end{flushright}
\vspace{.3cm}
\begin{center}
\renewcommand{\thefootnote}{\fnsymbol{footnote}}
{\Large{\bf Instantons and Wormholes In Minkowski and (A)dS Spaces}}
\vskip1cm
{\large\bf Michael Gutperle$^1$\footnote{email:
gutperle@riemann.harvard.edu  }
and Wafic Sabra$^2$\footnote{email: ws00@aub.edu.lb}}
\renewcommand{\thefootnote}{\arabic{footnote}}
\setcounter{footnote}{0}
\vskip1cm
{\small
$^1$Jefferson Physical Laboratory, Harvard University\\
Cambridge, MA 02138, USA\\
\vspace*{0.4cm}
$^2$ Center for Advanced Mathematical Sciences (CAMS)
and\\
Physics Department, American University of Beirut, Lebanon.\\}
\end{center}
%\vfill

\bigskip

\begin{center}
{\bf Abstract}
\end{center}
Instanton and wormhole solutions are constructed in a  $d$-dimensional
gravity theory with an axion-dilaton pair of scalar fields. We discuss the
cases of vanishing, positive and negative cosmological constant.  

\end{titlepage}

\baselineskip=20pt

\section{\protect\bigskip Introduction}

Wormholes may have relevance to many interesting questions in quantum
gravity. For instance, in the past it has been argued that they can play
role in the renormalisation of the coupling constants in nature, topological
fluctuations, quantum decoherence, the question of the vanishing of the
cosmological constant and creation of baby universes (see for example \cite
{stom, hawking, hawkingb, GRT,coleman,colemanb, Giddings:1988wv,
gids,Giddings:cg, Klebanov:1988eh, banks, giddingsa}).

Instanton solutions in supergravity theories can be responsible for
non-perturbative effects in string theories. For example the D-instanton
solution of type IIB supergravity found in \cite{Gibbons:1995vg} induces
higher derivative non-perturbative corrections of the IIB action \cite
{greengut}. Supergravity solutions corresponding to Euclidean wrapped branes
were found in \cite{Becker:1995kb,Becker:1999pb,Gutperle:2000sb, mspal}.

Axionic wormholes in four dimensions were first considered in \cite
{Giddings:cg}, where a system consisting of an axion, described by a rank
three antisymmetric tensor, coupled to gravity was considered. The case of
string theory was considered in \cite{giddingsa,rey} where in addition to
the axion one also includes a massless dilaton. The solutions are
characterized by an integration constant and instanton solutions are
extremal in the sense that the integration constant vanishes and some of the
(Euclidean) supersymmetry is preserved, whereas wormholes are non-extremal
and supersymmetry is broken \cite{Park:ep}.

In semi-classical quantum gravity, the action of an Euclidean wormhole
solution is related to the weight for the insertion of wormhole operators in
the path integral (or the rate of baby universe production). For stringy
wormholes it was found \cite{giddingsa,rey} that the existence of
non-singular solutions with finite action depended on the details of the
coupling of the dilaton to the axion. In particular for axion-dilaton
systems arising from string theory compactification there are no finite
action wormholes. However, it was argued that instantons could still
contribute when dilaton acquires mass due to supersymmetry breaking.
Generalization of axionic instantons to the case of a positive cosmological
constant was later obtained in \cite{Myers:ex, Myers:sp}.

In this paper we study instantons and wormholes for an axion-dilaton system
in $d$ dimensions and obtain a formula for critical values for couplings
below which non-singular solutions do exist. We will first consider the
Minkowski case and thus obtain generalizations of the known results in four
dimensions to an arbitrary $d$-dimensional space. Our main result is the
study of instantons and wormholes in de Sitter and Anti-de Sitter spaces.

\section{ Axion-Dilaton gravity}

We start our analysis by considering the theory of $d$-dimensional gravity
coupled to an axion-dilaton system. We will find generalizations of the
results of \cite{giddingsa, rey} to arbitrary dimensions. For Minkowski
signature, the action we consider is given by 
\begin{equation}
S_{m}=\int d^{d}x\sqrt{-g}\left( R-{\frac{1}{2}}\partial _{\mu }\phi
\partial ^{\mu }\phi -{\frac{1}{2}}e^{b\phi }\partial _{\mu }\chi \partial
^{\mu }\chi -V(\phi )\right) .
\end{equation}
The potential $V$ will not depend on the axion $\chi $ since we assume an
exact shift symmetry $\chi \rightarrow \chi +\epsilon $. The value of $b$ is
determined by the particular theory one considers\footnote{%
For example, in ten dimensional type IIB string theory one has $b=2$.}. In $d
$ dimensions the axion $\chi $ can be dualized to a $d-1$ form field
strength $F_{d-1}=dC_{d-2}$ via $d\chi =e^{-b\phi }\ast F_{d-1}$. The
dualized action takes the form

\begin{equation}
S_{m}^{\prime }=\int d^{d}x\sqrt{-g}\left( R-{\frac{1}{2}}\partial _{\mu
}\phi \partial ^{\mu }\phi -{\frac{e^{-b\phi }}{2(d-1)!}}F_{d-1}^{2}-V(\phi
)\right) .
\end{equation}
In this form the continuation to Euclidean signature is not problematic \cite
{Gibbons:1995vg, greengut, Cremmer:1998em}. However the dualization and
analytic continuation to Euclidean signature do not commute. This leads to
the fact that upon continuation to Euclidean space and dualization the
kinetic term for the axion $\chi $ changes sign 
\begin{equation}
S_{eucl}=\int d^{d}x\sqrt{g}\left( R-{\frac{1}{2}}\partial _{\mu }\phi
\partial ^{\mu }\phi +{\frac{1}{2}}e^{b\phi }\partial _{\mu }\chi \partial
^{\mu }\chi -V(\phi )\right) .  \label{bulac}
\end{equation}

In addition to the bulk term (\ref{bulac}) there is a boundary term in the
action which is important for the proper definition of the variational
principle and also for the calculation of the action \cite{Gibbons:1995vg,
greengut, giddingsa,rey}. The boundary term is given by 
\begin{equation}
S_{b}=\oint \left( e^{{b}\phi }\chi \partial _{n}\chi +K\right) ,
\label{bounda}
\end{equation}
where $K$ is the intrinsic curvature on the boundary. The Euclidean
equations of motion derived from (\ref{bulac}) are given by 
\begin{align}
\nabla _{\mu }(e^{b\phi }\partial ^{\mu }\chi )& =0, \\
\nabla ^{2}\phi +{\frac{b}{2}}e^{b\phi }(\partial \chi )^{2}-\frac{\partial
V(\phi )}{\partial \phi }& =0, \\
R_{\mu \nu }-{\frac{1}{2}}\partial _{\mu }\phi \partial _{\nu }\phi +{\frac{1%
}{2}}e^{b\phi }\partial _{\mu }\chi \partial _{\nu }\chi -{\frac{1}{d-2}}%
g_{\mu\nu}V(\phi )& =0.  \label{g}
\end{align}
In the following we will only consider a potential independent of $\phi $
corresponding to a cosmological constant $V=\Lambda $. We are also
interested in the most symmetric instanton and wormhole solutions and so we
shall consider the following $O(d)$ invariant ansatz for the metric 
\begin{equation}
ds^{2}=dr^{2}+a(r)^{2}d\Omega _{d-1}^{2},  \label{ansamet}
\end{equation}
where $d\Omega _{d-1}^{2}$ is the metric of the $\left( d-1\right) $ round
sphere $S_{d-1}$. In addition we demand that the dilaton $\phi $ and axion $%
\chi $ depend only on the radial coordinate $r$.

One could imagine an ansatz where the sphere metric $d\Omega_{d-1}^{2}$ is
replaced by the metric on flat space $M_{d-1}$ or the metric on hyperbolic
space $H_{d-1}$. However such a slicing would lead to solutions with
infinite action coming from (\ref{bounda}), because the (unit) volume of $%
M_{d-1}$ and $H_{d-1}$ is infinite and one integrates over the volume in (%
\ref{bounda}).

\ For the ansatz (\ref{ansamet}), the equation of motion for the axion gives 
\begin{equation}
\partial _{r}\left( e^{b\phi }a^{d-1}\partial _{r}\chi \right) =0,
\end{equation}
which implies a first-order differential equation for the axion field given
by 
\begin{equation}
\partial _{r}\chi =\frac{q}{a^{d-1}}e^{-b\phi }.  \label{axionsol}
\end{equation}
Upon using (\ref{axionsol}), we obtain from the gravitational equations of
motion (\ref{g}) 
\begin{align}
-(d-1)\frac{\partial _{r}^{2}a}{a}-{\frac{1}{2}}\left( \partial _{r}\phi
\right) ^{2}+\frac{q^{2}}{2a^{2d-2}}e^{-b\phi }-{\frac{1}{d-2}}\Lambda & =0,
\notag \\
(d-2)(1-(\partial _{r}a)^{2})-a\partial _{r}^{2}a-{\frac{1}{d-2}}%
a^{2}\Lambda & =0,  \label{graveq}
\end{align}
where we have used 
\begin{eqnarray}
R_{rr} &=&-(d-1)\frac{\partial _{r}^{2}a}{a},  \notag \\
R_{ij} &=&\left[ (d-2)(1-(\partial _{r}a)^{2})-a\partial _{r}^{2}a\right]
\delta _{ij}.
\end{eqnarray}
Also, upon using (\ref{axionsol}) and our metric ansatz, the dilaton
equation of motion becomes 
\begin{equation}
\partial _{r}^{2}\phi +\left( d-1\right) \frac{\partial _{r}a}{a}\partial
_{r}\phi +\frac{bq^{2}}{2a^{2d-2}}e^{-b\phi }=0.  \label{phieq}
\end{equation}
It follows from (\ref{phieq}) that there exists an integral 
\begin{equation}
\left( \partial _{r}\phi \right) ^{2}-\frac{q^{2}}{a^{2d-2}}e^{-b\phi }-%
\frac{c}{a^{2d-2}}=0,  \label{phieqb}
\end{equation}
where $c$ is a constant of integration. Using this integral, the
gravitational equations of motion (\ref{graveq}) give 
\begin{equation}
1-(\partial _{r}a)^{2}+\frac{c}{2(d-1)(d-2)a^{2d-4}}-{\frac{\Lambda }{%
(d-1)(d-2)}}a^{2}=0.  \label{greq}
\end{equation}
Equation (\ref{greq}) can be integrated to find $a(r)$ and (\ref{phieqb})
can then be solved to find $\phi (r)$. The detailed properties of these
solutions will depend on the choice of $\Lambda $ and $c$. For $c=0$ one
finds an `extremal' instanton whereas for $c\neq 0$ one finds `non-extremal'
wormhole solution. The cases of $\Lambda =0,\Lambda <0$ and $\Lambda >0$
correspond to (asymptotically) flat, anti-de Sitter and de Sitter space
respectively. All the possible cases will be discussed in the following.

\section{Instantons}

For $c=0,$ (\ref{greq}) can be solved and one finds the three solutions
corresponding to flat, de Sitter and anti-de Sitter space respectively. 
\begin{eqnarray}
\Lambda =0: &\quad &a^{2}(r)=r^{2}, \\
\Lambda >0: &\quad &a^{2}(r)={\frac{(d-1)(d-2)}{\Lambda }}\sin ^{2}\Big(%
\sqrt{\frac{\Lambda }{(d-1)(d-2)}}\;r\Big), \\
\Lambda <0: &\quad &a^{2}(r)={\frac{(d-1)(d-2)}{|\Lambda |}}\sinh ^{2}\Big(%
\sqrt{\frac{|\Lambda |}{(d-1)(d-2)}}\;r\Big).
\end{eqnarray}
If $c=0$ in (\ref{phieqb}), the solution of the dilaton is of the form 
\begin{equation}
e^{{\frac{b}{2}}\phi }=\mathrm{const}-{\frac{|bq|}{2}}\int {\frac{dr}{%
a(r)^{d-1}}}.  \label{instsol}
\end{equation}

\subsection{Flat space}

For $\Lambda =0$, the metric (in the Einstein frame) is flat space. The
dilaton (\ref{instsol}) is then given by 
\begin{equation}
e^{{\frac{b}{2}}\phi }=e^{{\frac{b}{2}}\phi _{\infty }}+\frac{|bq|}{2(d-2)}%
\frac{1}{r^{d-2}},
\end{equation}
where $\phi _{\infty }$ is the value of the dilaton at infinity. The fact
that the equations of motion are equivalent to a first order equation is a
hint that this solution is related to a Killing spinor equation. This
implies that in a supergravity theory these solutions are BPS and preserve
some supersymmetry. We also note that the dilaton diverges as $r\rightarrow
0 $. However the action (\ref{bounda}) of the instanton is finite and solely
given by a contribution from infinity 
\begin{equation}
S_{inst}={Vol(S_{d-1}) 2|q|\over |b|}e^{-{\frac{b}{2}}\phi _{\infty }}.  \label{instac}
\end{equation}
Such instantons are responsible for non-perturbative effects. The broken
supersymmetries are related to fermionic zero modes and integration over
fermionic zero modes induces higher dimensional terms in the effective
action. Such terms are weighted by the instanton action (\ref{instac}).
Often such effects can be attributed to Euclidean wrapped branes or
D-instantons \cite{Becker:1995kb,Becker:1999pb,Gutperle:2000sb, mspal}.

\subsection{Anti-de Sitter space}

In this case the dilaton solution for the instanton in AdS space is given by 
\begin{eqnarray}
e^{{\frac{b}{2}}\phi } &=&\mathrm{const}-{\frac{|bq|}{2}}\int {\frac{dr}{%
a(r)^{d-1}}}  \notag \\
&=&\mathrm{const}-{\frac{|bq|}{2}}\left( {\frac{|\Lambda |}{(d-1)(d-2)}}%
\right) ^{{\frac{d-1}{2}}}\int {\frac{dr}{\sinh ^{d-1}\Big(\sqrt{\frac{%
|\Lambda |}{(d-1)(d-2)}}\;r\Big)}}.
\end{eqnarray}
The behavior of the instanton is very similar to the one of flat space, in
particular the action of the instanton is again given by a boundary term 
\begin{equation}
S_{inst}={Vol(S_{d-1}) 2 |q|\over |b|}e^{-{\frac{b}{2}}\phi _{\infty }}.  \label{instacb}
\end{equation}
In the calculation of the action of the above Euclidean gravitational
instantons, one has to include the contribution of the gravitational action
which diverges because of the infinite volume of AdS. To render the action
finite one needs to employ the counterterms subtraction method which was
proposed for computing the boundary stress tensor associated with a
gravitating system \cite{Kraus,ejr,Nojiri:1999jj,Nojiri:2000kh}. The appearance of the logarithmically
divergent contributions has a physical meaning\footnote{%
The counterterms subtraction method of \cite{Kraus} was employed in \cite
{ejr} and explicit calculations have been performed for the gravitational
action for our metric in the dimensions $d=3,4,5,6,7$. Logarithmically
divergent contributions have been obtained for $d=3,5,7.$%
\par
\bigskip}. They are related to the conformal anomaly of the dual field
theory living on the boundary. In some sense the AdS/CFT correspondence
relates the infinite volume singularity in the AdS to the UV divergence in
the CFT. This regularization is independent of the presence of an instanton
and the instanton ($q$ dependent) part of the action is given by (\ref
{instacb}).

\subsection{de Sitter space}

\label{dsinst} In this case the dilaton solution for the instanton in dS
space is given by 
\begin{eqnarray}
e^{{\frac{b}{2}}\phi } &=&\mathrm{const}-{\frac{|bq|}{2}}\int {\frac{dr}{%
a(r)^{d-1}}}  \notag \\
&=&\mathrm{const}-{\frac{|bq|}{2}}\left( {\frac{\Lambda }{(d-1)(d-2)}}%
\right) ^{{\frac{d-1}{2}}}\int {\frac{dr}{\sin ^{d-1}\Big(\sqrt{\frac{%
\Lambda }{(d-1)(d-2)}}\;r\Big)}}.
\end{eqnarray}
Since Euclidean $d$-dimensional de Sitter space is simply the $d$%
-dimensional sphere the range of $r$ is $r\in \lbrack 0,\pi \sqrt{\frac{%
(d-1)(d-2)}{\Lambda }}]$. The integral on the right hand side ranges between
-$\infty $ and +$\infty $. Since $e^{{\frac{b}{2}}\phi }$ ranges between 0
and +$\infty $ , the dilaton must become singular and therefore instantons
do not exist in de Sitter space.

\section{Wormholes}

If $c\neq 0$ the equations become more complicated. As we shall see
solutions in this case correspond to wormholes. This means that the metric
factor $a(r)$ reaches a minimal value $a_{0}>0$ where $\partial _{r}a(r)=0$,
defining the neck of the wormhole. Here

\begin{equation}
\partial _{r}a=\pm \sqrt{1+\frac{c}{2(d-1)(d-2)a^{2d-4}}-{\frac{\Lambda }{%
(d-1)(d-2)}}a^{2}}.  \label{greqd}
\end{equation}
Hence there is a sphere of non-zero minimal size. 
The dilaton can be determined from (\ref{phieqb}) and we
get 
\begin{equation}
\int {\frac{d\phi }{\sqrt{q^{2}e^{-b\phi }+c}}}=\pm \int {\frac{dr}{a^{d-1}}}%
.
\end{equation}
In the following we will solve these equations for all possible values of $%
\Lambda $ and $c$.

 An Euclidean wormhole corresponds to a tunneling event between two
asymptotic spaces.    
 Alternatively, one can cut the wormhole in half along the minimal 
 sphere. Such an Euclidean configuration provides the bounce for the 
 creation of a baby universe. The continuation to Minkowski signature 
 is consistent since the momentum components of the fields normal to 
 the minimal sphere vanish.\footnote{For the axion it is convenient to 
   dualize the to a $d-1$ form whose indices are tangent to the 
   minimal sphere.}  The Minkowskian evolution of the baby universe
 will be of FRW type.

\subsection{Flat space}

For $\Lambda =0,$ the space will be asymptotically flat. Recall from (\ref
{greq}) that 
\begin{equation}
1-(a^{\prime })^{2}+{\frac{c}{2(d-1)(d-2)}}{\frac{1}{a^{2d-4}}}=0.
\end{equation}
If $c<0,$ one has a minimal size sphere at $a_{0}=\left( {\frac{2(d-1)(d-2)}{%
|c|}}\right) ^{\frac{1}{2d-4}},$ where $\partial _{r}a=0.$ This is the neck
of the wormhole, connecting two asymptotically flat regions, located at $%
r\rightarrow \pm \infty $. Integrating the $\phi $ equation (\ref{phieqb})
gives

\begin{equation}
\frac{1}{|c|^{1/2}|b|}\int {\frac{d\tilde{\phi}}{\sqrt{e^{-\tilde{\phi}}-1}}}%
=\pm \int \frac{da}{a^{d-1}\sqrt{1-{\frac{|c|}{2(d-1)(d-2)}}{\frac{1}{%
a^{2d-4}}}}},  \label{wormeq}
\end{equation}
where we have defined 
\begin{equation}
e^{b\phi }={\frac{q^{2}}{|c|}}e^{\tilde{\phi}}.
\end{equation}
Equation (\ref{wormeq}) can be easily integrated. Using the fact that $%
a\rightarrow \infty $ corresponds to $\phi \rightarrow \phi _{\infty }$ one
finds the following solution 
\begin{align}
& \arcsin \left( \sqrt{\frac{|c|}{q^{2}}}e^{{\frac{b}{2}}\phi (r)}\right)
-\arcsin \left( \sqrt{\frac{|c|}{q^{2}}}e^{{\frac{b}{2}}\phi _{\infty
}}\right)  \notag \\
& =\mp |b|\sqrt{\frac{d-1}{2(d-2)}}\arcsin {\small {\ }}\left( {\small \sqrt{%
\frac{|c|}{2(d-1)(d-2)}}{\frac{1}{a(r)^{d-2}}}}\right) .  \label{wormrel}
\end{align}
As one approaches the neck of the wormhole ($a\rightarrow a_{0})$, the
argument of the $\arcsin $ on right hand side of (\ref{wormrel}) becomes
one. The dilaton should be regular on the neck. This implies that there is a
'critical value' $b_{c}$ of the dilaton coupling and for non-singular
solutions we must have\footnote{%
Note that in \cite{giddingsa}, the derivative of the dilaton was assumed to
vanish at the throat of the wormhole. Moreover, our dilaton solution for the
case of $d=4,$ differs from that given in \cite{rey} though we agree on the
critical value $b_{c}.$} 
\begin{equation}
|b|<b_{c}=\sqrt{\frac{2(d-2)}{d-1}.}  \label{bcrit}
\end{equation}
Note that the integration constant $c$ does not appear in (\ref{bcrit}). The
Euclidean action of the wormhole is still given by a boundary term only,
however in contrast to the instanton case the region of the neck of the
wormhole does contribute. 
\begin{equation}
S_{eucl}={2\over |b|} Vol(S_{d-1})\left(\sqrt{q^2 e^{-b\phi_\infty}+c}-  \sqrt{q^2 
e^{-b\phi_0}+c}\right) .  \label{whact}
\end{equation}
In order to have a positive wormhole action (\ref{whact}) and a smooth limit
to the instanton solution as $|c|\rightarrow 0$ we have to choose the plus
sign in (\ref{wormrel}).

Here $\phi _{0}$ is the value of the dilaton at the neck and is given by 
\begin{equation}
e^{{\frac{b}{2}}\phi (r_{0})}=\frac{|q|}{|c|^{\frac{1}{2}}}\sin \left(
\arcsin \left( \sqrt{\frac{|c|}{q^{2}}}e^{{\frac{b}{2}}\phi _{\infty
}}\right) +|b|\frac{\pi }{2}\sqrt{\frac{d-1}{2(d-2)}}\right) .
\end{equation}
The finiteness of the wormhole action depends on the value of $b$. If $%
|b|>b_{c}$ the Euclidean action is infinite and the wormholes are completely
suppressed in the semi-classical approximation. The above analysis
constitutes the generalization of the results known in the literature for
four dimensions to $d$ dimensions.

\subsection{Anti de Sitter space}

\label{adssec}

For $c<0,$ the existence of the integral (\ref{phieqb}) implies the relation 
\begin{equation}
{\frac{1}{\sqrt{|c|}|b|}}\int {\frac{d\tilde{\phi}}{\sqrt{e^{-\tilde{\phi}}-1%
}}}=\pm \int {\frac{da}{a^{d-1}\sqrt{1-\frac{|c|}{2(d-1)(d-2)}{\frac{1}{%
a^{2d-4}}}+{\frac{|\Lambda |}{(d-1)(d-2)}}a^{2}}}}.  \label{adsint}
\end{equation}
The above equation can be easily integrated in terms of elementary functions
for the case of $d=3.$ In this case we obtain

\begin{align}
& \arcsin \left( \sqrt{\frac{|c|}{q^{2}}}e^{{\frac{b}{2}}\phi (r)}\right)
-\arcsin \left( \sqrt{\frac{|c|}{q^{2}}}e^{{\frac{b}{2}}\phi _{\infty
}}\right)   \notag \\
{}& ={}\pm \frac{|b|}{2}\left( \arcsin \left( {\frac{-{\frac{|c|}{2}+a}^{2}{%
(r)}}{{a}^{2}{(r)}\sqrt{1+\frac{|c\Lambda |}{2}}}}\right) -\arcsin \left( {%
\frac{1}{\sqrt{1+\frac{|c\Lambda |}{2}}}}\right) \right) .  \label{adswhs}
\end{align}
This wormhole solutions for certain dilaton couplings would not be finite if
one turns on $\Lambda $. The neck is at 
\begin{equation}
a_{0}^{2}=\frac{\sqrt{1+\frac{{|\Lambda c|}}{2}}-1}{|{\Lambda |}}.
\end{equation}
At the neck we obtain 
\begin{align}
& \arcsin \left( \sqrt{\frac{|c|}{q^{2}}}e^{{\frac{b}{2}}\phi _{0}}\right)
-\arcsin \left( \sqrt{\frac{|c|}{q^{2}}}e^{{\frac{b}{2}}\phi _{\infty
}}\right)   \notag \\
{}& ={}\mp \frac{|b|}{2}\left( \frac{\pi }{2}+\arcsin \left( {\frac{1}{\sqrt{%
1+\frac{|c\Lambda |}{2}}}}\right) \right) .
\end{align}
In this case the critical value of $|b|<b_{c}$ is given by 
\begin{equation}
b_{c}=2\left[ 1+\frac{2}{\pi }\arcsin \left( {\frac{1}{\sqrt{1+\frac{%
|c\Lambda |}{2}}}}\right) \right] ^{-1}.
\end{equation}
Hence setting $\Lambda \rightarrow 0$ one gets $b_{c}=1$ which is the
correct flat space value (See (\ref{bcrit}) for $d=3$).

The choice of the negative sign in (\ref{adswhs}) leads to a positive action
and smooth limit to the anti de Sitter instanton solution\footnote{%
For the instanton solution in anti-de Sitter three dimensional space we have 
\begin{equation*}
e^{{\frac{b}{2}}\phi (r)}-e^{{\frac{b}{2}}\phi _{\infty }}=\frac{|bq|}{2}%
\sqrt{\frac{|\Lambda |}{2}}\left( \frac{\cosh \sqrt{\frac{|\Lambda |}{2}}r}{%
\sinh \sqrt{\frac{|\Lambda |}{2}}r}-1\right)
\end{equation*}
}.
 The range of $a(r)$ is $[a_0,\infty]$ and as $r\to \infty$ the
  metric approaches the euclidean AdS metric, hence the wormhole
  connects two asymptotically AdS spaces.

Note however that for non-zero $\Lambda $ the value of $b_{c}$ decreases.
These wormhole solutions for certain dilaton couplings would not be finite
if one turns on a negative cosmological constant $\Lambda $. For $d=4,5$ the
integral on the right hand side of (\ref{adsint}) can still be solved in
terms of elliptic integrals, however the results are not very illuminating,
the general feature that there is a critical value of $b_{c}$ which is
smaller than the flat space value persist. For $d>5$ one can only solve the
integrals numerically.

\subsection{de Sitter space}

\label{dssec}

In this section we will consider wormhole solutions for de Sitter spaces,
i.e., for the case of positive cosmological constant. In this case the
equation of the metric takes the form (for $\Lambda >0)$ 
\begin{equation}
1-(\partial _{r}a)^{2}+{\frac{c}{2(d-1)(d-2)}}{\frac{1}{a^{2d-4}}-\frac{%
\Lambda }{(d-1)(d-2)}}a^{2}=0.  \label{metqa}
\end{equation}
The instanton for which $c=0$ was discussed in section \ref{dsinst}. For $%
c\neq 0$, the dilaton is determined by the integral 
\begin{equation}
\int \frac{d\phi }{\sqrt{q^{2}e^{-b\phi }+c}}=\pm \int \frac{da}{a^{d-1}%
\sqrt{1+\frac{c}{2(d-1)(d-2)a^{2d-4}}-{\frac{\Lambda }{(d-1)(d-2)}}a^{2}}}.
\end{equation}
In what follows we will consider the three dimensional case for simplicity.
It is clear that the two cases $c>0$ and $c<0$ are qualitatively different
and will discuss both of them in turn. For $c<0$, and for the case of $d=3$,
(\ref{metqa}) can have either two or no values of $a$ where $\partial
_{r}a=0 $, if 
\begin{equation}
0<|c\Lambda |<2,
\end{equation}
then we have two real zeros given by 
\begin{equation}
a_{\pm }^{2}=\frac{1\pm \sqrt{1-\frac{{|\Lambda c|}}{2}}}{{\Lambda }}.
\end{equation}
The solution for the dilaton is given by 
\begin{equation}
\arcsin \sqrt{\frac{|c|}{q^{2}}}e^{{\frac{b}{2}}\phi (r)}-\arcsin \sqrt{%
\frac{|c|}{q^{2}}}e^{{\frac{b}{2}}\phi _{-}}=\pm \frac{|b|}{2}\left( \arcsin
\left( {\frac{2a^{2}(r)-|c|}{2a^{2}(r)\sqrt{1-\frac{|c\Lambda |}{2}}}}%
\right) +{\frac{\pi }{2}}\right) ,
\end{equation}
where $\phi \rightarrow \phi _{\pm }$ as $a\rightarrow a_{\pm }$. Note that
a non-singular dilaton configuration is only possible if the dilaton
coupling $b$ satisfies 
\begin{equation}
|b|<b_{c}=1.
\end{equation}
Here we have only displayed the results for $d=3$. For $d=4$ and $d=5,$ the
relevant integrals can be solved in terms of elliptic integrals, however the
results are not very illuminating. For $d>5,$ the integrals can only be
solved numerically.

 The range of $a(r)$ is $a\in [a_-,a_+]$ and the topology of the
  wormhole is the one of a cylinder. A
 wormhole with opposite axion charge could be glued onto the
first one. As pointed out in \cite{giddingsa}, the dilaton makes this
impossible since it would have to have vanishing derivative at both necks in
order to be continuous. This is only possible if the dilaton becomes
singular.

Now for $c>0,$ there is only one value of $a$ where $\partial _{r}a=0$, 
\begin{equation}
a_{0}^{2}=\frac{1+\sqrt{1+\frac{c\Lambda }{2}}}{{\Lambda }},
\end{equation}
hence the metric function has the range $a\in \lbrack 0,a_{0}]$. The
solution for the dilaton equation in this case reads 
\begin{equation}
\mathrm{arcsinh}\left( \sqrt{\frac{c}{q^{2}}}e^{{\frac{b}{2}}\phi
(r)}\right) -\mathrm{arcsinh}\left( \sqrt{\frac{c}{q^{2}}}e^{{\frac{b}{2}}%
\phi _{0}}\right) =\mp \frac{|b|}{2}\mathrm{arcosh}\left( {\frac{2a^{2}(r)+c%
}{2a^{2}(r)\sqrt{1+{\frac{c\Lambda }{2}}}}}\right) .
\end{equation}
where $\phi _{0}$ is the value of the dilaton at $a=a_{0}$. Note however
that this solution contains a curvature singularity at $a=0$. At this point $%
a(r)\sim r^{1/(d-1)}$ as $r\rightarrow 0$ and it is easy to show the Ricci
scalar diverges. Since only non-singular solutions are admissible as
semiclassical saddle points, the solution with $c>0$ cannot be used as an
Euclidean wormhole.

\section{Axionic Solutions}

The case of $b=0$ is special since then there is no coupling of the axion to
the dilaton and the dilaton can be consistently set to zero. The axion is
still given by the equation 
\begin{equation}
\partial _{r}\chi =\frac{q}{a^{d-1}}.  \label{axeqa}
\end{equation}
The role of the integration constant $c$ is now simply played by $-q^{2}$.
The gravitational equations are equivalent to 
\begin{equation}
(\partial _{r}a)^{2}=1-{\frac{1}{2}}\frac{q^{2}}{(d-1)(d-2)a^{2d-4}}-\frac{%
\Lambda a^{2}}{(d-1)(d-2)}.  \label{ddd}
\end{equation}
Here it is useful to perform a change of variable $r^{\prime }=a(r),$ and we
obtain for the metric 
\begin{equation}
ds^{2}=\frac{1}{1-{\frac{1}{2}}\frac{q^{2}}{(d-1)(d-2)r^{2d-4}}-\frac{%
\Lambda r^{2}}{(d-1)(d-2)}}dr^{2}+r^{2}d\Omega _{d-1}^{2}.
\end{equation}
In these coordinates (\ref{axeqa}) can be solved 
\begin{equation}
\chi =\mathrm{const}\pm \int dr\frac{q}{r^{d-1}}\frac{1}{\sqrt{1-{\frac{1}{2}%
}\frac{q^{2}}{(d-1)(d-2)r^{2d-4}}-\frac{\Lambda r^{2}}{(d-1)(d-2)}}}.
\label{chiint}
\end{equation}
The integral (\ref{chiint}) is very similar to the ones discussed in the
previous section. For $\Lambda =0$ the solution is 
\begin{equation}
\chi -\chi _{\infty }=\mp \sqrt{{\frac{2(d-1)}{(d-2)}}}\arcsin \sqrt{{\frac{1%
}{2}}\frac{q^{2}}{(d-1)(d-2)r^{2d-4}}}.
\end{equation}
For $\Lambda \neq $ $0$, explicit formulae can be obtained for $d=3,4,5$ as
discussed in section \ref{adssec} and \ref{dssec}. Again only the case of $%
d=3$ has a simple realization in terms of elementary functions. For example
for $\Lambda <0$ and $d=3$ one finds 
\begin{equation}
\chi -\chi _{\infty }=\pm \left( \arcsin \left( {\frac{-{\frac{q^{2}}{2}+r}%
^{2}}{{r}^{2}\sqrt{1+\frac{|q^{2}\Lambda |}{2}}}}\right) -\arcsin \left( {%
\frac{1}{\sqrt{1+\frac{|q^{2}\Lambda |}{2}}}}\right) \right) .
\end{equation}

\section{Discussion}

In this paper we have discussed instanton and wormhole solutions in
axion-dilaton gravity theories. Such theories are abundant in string
theories and supergravities in various dimensions. Instanton solutions exist
for flat and AdS spaces in arbitrary dimensions, however in dS space there
are no non-singular instanton solutions.

The existence of wormholes depends on the dimensionality of spacetime and
the dilaton coupling constant $b$. There is a critical value $b_{c}$ above
which the instanton solution becomes singular and has an infinite action.
Wormholes induce non-local interactions in the theory and there may be a
connection between wormholes and non-local string theories found in\cite
{Aharony:2001pa,Aharony:2001dp}. It would be interesting to see whether one
can find an example in string theory where $b$ is in the range where
non-singular wormholes exist.

In certain theories (e.g. in gauged supergravities), the dilaton can acquire
a non-trivial potential, 
\begin{equation}
S_{m}=\int d^{d}x\sqrt{-g}\left( R-{\frac{1}{2}}\partial _{\mu }\phi
\partial ^{\mu }\phi -{\frac{1}{2}}e^{b\phi }\partial _{\mu }\chi \partial
^{\mu }\chi -V(\phi )\right) .
\end{equation}
Using the ansatz for the metric given by (\ref{ansamet}), the dilaton
equation of motion is given by 
\begin{equation}
\partial _{r}^{2}\phi +\left( d-1\right) \frac{\partial _{r}a}{a}\partial
_{r}\phi +\frac{bq^{2}}{2a^{2d-2}}e^{-b\phi }-\frac{\partial V(\phi )}{%
\partial \phi }=0,  \label{phiec}
\end{equation}
and it is not possible to find a simple integral like (\ref{phieqb}) which
allowed for the decoupling of the metric and dilaton equations. Hence the
equations in these cases are considerably more complicated. However it may
be possible to find first order differential equations as in the case for
domain walls \cite{DeWolfe:1999cp,Skenderis:1999mm}. Also it might be
interesting to study these equations further and in particular investigate
whether it is possible to stabilize the dilaton and get wormhole solutions
which are more like axionic ones. We hope to report on these issues in a
future publication.

\bigskip

\textbf{Acknowledgments:}

\medskip

The work of M.G. was supported in part by DOE grant DE-FG02-91ER40655. M.G.
gratefully acknowledges the hospitality of the Stanford Theory Group during
the final stages of this work. We are grateful to U. Theis, S. Vandoren and K.
Behrndt for pointing out an error in a previous version.


\begin{thebibliography}{99}
\bibitem{stom}  A.~Strominger, ``Vacuum Topology And Incoherence In Quantum
Gravity,'' Phys.\ Rev.\ Lett.\ \textbf{52} (1984) 1733. 
%%CITATION = PRLTA,52,1733;%%

\bibitem{hawking}  S.~W.~Hawking, ``Quantum Coherence Down The Wormhole,''
Phys.\ Lett.\ B \textbf{195} (1987) 337. %%CITATION = PHLTA,B195,337;%%

\bibitem{hawkingb}  S.~W.~Hawking, ``Wormholes In Space-Time,'' Phys.\ Rev.\
D \textbf{37} (1988) 904. %%CITATION = PHRVA,D37,904;%%

\bibitem{GRT}  G.~V.~Lavrelashvili, V.~A.~Rubakov and P.~G.~Tinyakov,
``Particle Creation And Destruction Of Quantum Coherence By Topological
Change,'' Nucl.\ Phys.\ B \textbf{299} (1988) 757. 
%%CITATION = NUPHA,B299,757;%%

\bibitem{coleman}  S.~R.~Coleman, ``Black Holes As Red Herrings: Topological
Fluctuations And The Loss Of Quantum Coherence,'' Nucl.\ Phys.\ B \textbf{307%
} (1988) 867. %%CITATION = NUPHA,B307,867;%%

\bibitem{colemanb}  S.~R.~Coleman, ``Why There Is Nothing Rather Than
Something: A Theory Of The Cosmological Constant,'' Nucl.\ Phys.\ B \textbf{%
310} (1988) 643. %%CITATION = NUPHA,B310,643;%%

\bibitem{Giddings:1988wv}  S.~B.~Giddings and A.~Strominger, ``Baby
Universes, Third Quantization And The Cosmological Constant,'' Nucl.\ Phys.\
B \textbf{321} (1989) 481. %%CITATION = NUPHA,B321,481;%%

\bibitem{gids}  S.~B.~Giddings and A.~Strominger, ``Loss Of Incoherence And
Determination Of Coupling Constants In Quantum Gravity,'' Nucl.\ Phys.\ B 
\textbf{307} (1988) 854. %%CITATION = NUPHA,B307,854;%%

\bibitem{Giddings:cg}  S.~B.~Giddings and A.~Strominger, ``Axion Induced
Topology Change In Quantum Gravity And String Theory,'' Nucl.\ Phys.\ B 
\textbf{306} (1988) 890. %%CITATION = NUPHA,B306,890;%%

\bibitem{Klebanov:1988eh}  I.~R.~Klebanov, L.~Susskind and T.~Banks,
``Wormholes And The Cosmological Constant,'' Nucl.\ Phys.\ B \textbf{317}
(1989) 665. %%CITATION = NUPHA,B317,665;%%

\bibitem{banks}  T.~Banks, ``Prolegomena To A Theory Of Bifurcating
Universes: A Nonlocal Solution To The Cosmological Constant Problem Or
Little Lambda Goes Back To The Future,'' Nucl.\ Phys.\ B \textbf{309} (1988)
493. %%CITATION = NUPHA,B309,493;%%

\bibitem{giddingsa}  S.~B.~Giddings and A.~Strominger, ``String Wormholes,''
Phys.\ Lett.\ B \textbf{230} (1989) 46. %%CITATION = PHLTA,B230,46;%%
%
%

\bibitem{Gibbons:1995vg}  G.~W.~Gibbons, M.~B.~Green and M.~J.~Perry,
``Instantons and Seven-Branes in Type IIB Superstring Theory,'' Phys.\
Lett.\ B \textbf{370} (1996) 37 [arXiv:hep-th/9511080]. 
%%CITATION = HEP-TH 9511080;%%

\bibitem{greengut}  M.~B.~Green and M.~Gutperle, ``Effects of
D-instantons,'' Nucl.\ Phys.\ B \textbf{498} (1997) 195 [hep-th/9701093]. 
%%CITATION = HEP-TH 9701093;%%
%
%

\bibitem{Gutperle:2000sb}  M.~Gutperle and M.~Spalinski, ``Supergravity
instantons and the universal hypermultiplet,'' JHEP \textbf{0006} (2000) 037
[arXiv:hep-th/0005068]. %%CITATION = HEP-TH 0005068;%%

\bibitem{mspal}  M.~Gutperle and M.~Spalinski, ``Supergravity instantons for
N = 2 hypermultiplets,'' Nucl.\ Phys.\ B \textbf{598} (2001) 509
[hep-th/0010192]. %%CITATION = HEP-TH 0010192;%%
%
%

\bibitem{Becker:1999pb}  K.~Becker and M.~Becker, ``Instanton action for
type II hypermultiplets,'' Nucl.\ Phys.\ B \textbf{551} (1999) 102
[arXiv:hep-th/9901126]. %%CITATION = HEP-TH 9901126;%%

\bibitem{Becker:1995kb}  K.~Becker, M.~Becker and A.~Strominger,
``Five-branes, membranes and nonperturbative string theory,'' Nucl.\ Phys.\
B \textbf{456} (1995) 130 [arXiv:hep-th/9507158]. 
%%CITATION = HEP-TH 9507158;%%

\bibitem{rey}  S.~Rey, ``The Confining Phase of Superstrings and Axionic
Strings,'' Phys.\ Rev.\ D \textbf{43} (1991) 526.

%%CITATION = PHRVA,D43,526;%%
%
%

\bibitem{Park:ep}  Y.~Park, M.~Srednicki and A.~Strominger, ``Wormhole
Induced Supersymmetry Breaking In String Theory,'' Phys.\ Lett.\ B \textbf{%
244} (1990) 393. %%CITATION = PHLTA,B244,393;%%

\bibitem{Myers:ex}  R.~C.~Myers, ``Baby Universes In Higher Dimensional
Theories,'' Nucl.\ Phys.\ B \textbf{323} (1989) 225. 
%%CITATION = NUPHA,B323,225;%%

\bibitem{Myers:sp}  R.~C.~Myers, ``New Axionic Instantons In Quantum
Gravity,'' Phys.\ Rev.\ D \textbf{38} (1988) 1327. 
%%CITATION = PHRVA,D38,1327;%%

\bibitem{Cremmer:1998em}  E.~Cremmer, I.~V.~Lavrinenko, H.~Lu, C.~N.~Pope,
K.~S.~Stelle and T.~A.~Tran, ``Euclidean-signature supergravities, dualities
and instantons,'' Nucl.\ Phys.\ B \textbf{534} (1998) 40
[arXiv:hep-th/9803259]. %%CITATION = HEP-TH 9803259;%%

\bibitem{Kraus}  V. Balasubramanian and P. Kraus, ``A Stress Tensor for
Anti-de Sitter Gravity,'' Commun.Math.Phys. \textbf{208} (1999) 413
[arXiv:hep-th/9902121].%%CITATION = HEP-TH 9906127;%%
%


\bibitem{ejr}  R.~Emparan, C.~V.~Johnson and R.~C.~Myers, ``Surface terms as
counterterms in the AdS/CFT correspondence,'' Phys.\ Rev.\ D \textbf{60}
(1999) 104001 [arXiv:hep-th/9903238]. %%CITATION = HEP-TH 9903238;%%

\bibitem{Nojiri:1999jj}
S.~Nojiri, S.~D.~Odintsov, S.~Ogushi, A.~Sugamoto and M.~Yamamoto,
``Axion-dilatonic conformal anomaly from AdS/CFT correspondence,''
Phys.\ Lett.\ B {\bf 465} (1999) 128
[arXiv:hep-th/9908066].
%%CITATION = HEP-TH 9908066;%%


\bibitem{Nojiri:2000kh}
S.~Nojiri, S.~D.~Odintsov and S.~Ogushi,
``Finite action in d5 gauged supergravity and dilatonic conformal anomaly
  for dual quantum field theory,''
Phys.\ Rev.\ D {\bf 62} (2000) 124002
[arXiv:hep-th/0001122].
%%CITATION = HEP-TH 0001122;%%

\bibitem{Aharony:2001pa}  O.~Aharony, M.~Berkooz and E.~Silverstein,
``Multiple-trace operators and non-local string theories,'' JHEP \textbf{0108%
} (2001) 006 [arXiv:hep-th/0105309]. %%CITATION = HEP-TH 0105309;%%

\bibitem{Aharony:2001dp}  O.~Aharony, M.~Berkooz and E.~Silverstein,
``Non-local string theories on AdS$_{3}\times $ S$^{3}$ and stable
non-supersymmetric backgrounds,'' arXiv:hep-th/0112178. 
%%CITATION = HEP-TH 0112178;%%

\bibitem{DeWolfe:1999cp}  O.~DeWolfe, D.~Z.~Freedman, S.~S.~Gubser and
A.~Karch, ``Modeling the fifth dimension with scalars and gravity,'' Phys.\
Rev.\ D \textbf{62} (2000) 046008 [arXiv:hep-th/9909134]. 
%%CITATION = HEP-TH 9909134;%%

\bibitem{Skenderis:1999mm}  K.~Skenderis and P.~K.~Townsend, ``Gravitational
stability and renormalization-group flow,'' Phys.\ Lett.\ B \textbf{468}
(1999) 46 [arXiv:hep-th/9909070]. %%CITATION = HEP-TH 9909070;%%
\end{thebibliography}
\end{document}